\documentclass[fleqn,twoside]{article}
\usepackage{espcrc2}

\usepackage{graphicx}
\usepackage[figuresright]{rotating}
\usepackage{amsmath}

\def\gsim{\raise0.3ex\hbox{$>$\kern-0.75em\raise-1.1ex\hbox{$\sim$}}}
\def\lsim{\raise0.3ex\hbox{$<$\kern-0.75em\raise-1.1ex\hbox{$\sim$}}}

\title{Superbradyons and some possible dark matter signatures}

\author{Luis Gonzalez-Mestres\address{LAPP, Universit\'e de Savoie, CNRS/IN2P3, B.P. 110, 74941 Annecy-le-Vieux Cedex, France}}

\begin{document}

\begin{abstract}
Superluminal preons (superbradyons) with a critical speed in vacuum much larger than the speed of light would, if they exist, play a fundamental role as constituents of the physical vacuum and of the conventional particles considered in standard theories. Then, standard Lorentz symmetry and quantum mechanics would not be ultimate fundamental properties of space-time and matter. If superbradyons are present as free particles in our Universe, they are expected to couple very weakly to "ordinary" matter, but they can spontaneously decay by emitting standard particles until they reach a speed equal or close to that of light. They would then form a cosmological sea where the relation between inertial and gravitational masses are expected to differ from conventional Physics. Superbradyons may be at the origin of cosmological and astrophysical phenomena usually associated to dark matter, dark energy and inflation. They can also be a source of conventional cosmic rays of all energies. In such a scenario, superbradyon spontaneous decays and similar interactions would be candidates to explain data on electron and positron abundances (PAMELA, ATIC, Fermi LAT, HESS, PPB-BETS) considered as possible dark matter signatures. Superbradyons emitting this radiation may even have comparatively small inertial masses, and kinetic energies not far away from those of the decay events. But in most cases, such "light" superbradyons could not be found at accelerator experiments. We comment on several basic physics and phenomenological issues in connection with the superbradyon hypothesis. 
\vspace{1pc}
\end{abstract}

\maketitle

\section{Introduction}

Lorentz symmetry violation (LSV) at the Planck scale \cite{Gonzalez-Mestres2008,Gonzalez-Mestres2009} is currently being tested by ultra-high energy cosmic-ray (UHECR) experiments \cite{AugerData1,AugerData2,HiresData1}. Data from the AUGER \cite{AUG} and HiRes \cite{Hires} collaborations possibly confirm the existence of the Greisen-Zatsepin-Kuzmin (GZK) cutoff \cite{GZK1,GZK2}. But in any case, several important LSV patterns and domains of parameters will remain allowed by these data \cite{Gonzalez-Mestres2009}. 

Furthermore, the AUGER Collaboration has recently reported \cite{AugerData3} a systematic inconsistency of available hadronic interaction models when attempting to simultaneously describe the observations of the $X_{max}$ parameter and the number $N_\mu $ of produced muons. Data on $X_{max}$ suggests UHECR masses to lie in the range between proton and iron, while $N_\mu $ data hint to heavier nuclei. 

As explained in our previous papers \cite{Gonzalez-Mestres2008,Gonzalez-Mestres2009}, the composition of the UHECR spectrum is a crucial issue. The experimental confirmation of the GZK cutoff would lead to significant bounds on LSV, but it would not by itself allow to exclude all possible LSV patterns and values of parameters. This is in particular the case for ultra-high energy (UHE) nuclei in the main Weak Doubly Special Relativity (WDSR, \cite{Gonzalez-Mestres2008,Gonzalez-Mestres2009,gonSL2}) models . 

In the family of models we suggested in 1997 \cite{gonSL2} using quadratically deformed relativistic kinematics (QDRK), UHECR masses heavier typically than $\approx ~30 $ times the proton mass would make compatible the observed flux suppression above $\approx ~ 3 ~ . ~10^{19}$ eV with a Lorentz symmetry violation of order $\approx ~1 $ at the Planck scale for nucleons. If the UHECR masses are heavier than $\approx ~10 $ times the proton mass, a LSV parameter $\approx ~1 $ at the Planck scale can still be allowed for quarks and gluons. 

The question of possible LSV does not only concern the kinematical and dynamical properties of standard particles, but also the very structure of matter at a deeper scale. At the Planck scale or at some other fundamental scale, the nature itself of elementary particles can drastically change. This may have important cosmological at inflation temperatures, as well as for dark matter and dark energy. 

Implications of recent experimental results for the superbradyon hypothesis \cite{gonSL1,gonSL2} have also been considered in our recent papers \cite{Gonzalez-Mestres2008,Gonzalez-Mestres2009}. No data allow to exclude the existence of such particles. Superbradyons would have a critical speed in vacuum $c_s$ much larger than the speed of light $c$, possibly corresponding to a superluminal Lorentz invariance. They may be the ultimate constituents of matter, and even obey a new mechanics as suggested below. 

Standard preon models \cite{preons} assumed that preons feel the same minkowskian space-time as quarks, leptons and gauge bosons, but there is no fundamental reason for this assumption. Preons were also assumed to carry the same kind of charges and quantum numbers as standard particles, but this is not necessarily the case if the gauge bosons are composite. 

Superbradyons do not follow the standard preonic pattern. They can obey a new Lorentz symmetry with their own critical speed in vacuum replacing the speed of light in the metrics, and not be directly coupled to standard electroweak and strong interactions. They may manifest themselves through Planck-scale phenomena not ruled out by present experimental data, or more directly through their presence in matter (including vacuum) and in the Universe producing observable effects.

Superbradyonic remnants of earlier cosmological eras can possibly exist nowadays as free particles, and be very difficult to detect if they are not directly coupled to standard interactions. Contrary to tachyons, they would have positive mass and energy, violate standard special relativity and spontaneously emit radiation in vacuum in the form of conventional particles. They may be sources of UHECR and of large-scale phenomena in our Universe. As a final result of "Cherenkov" decays, a cosmological sea of superbradyons with speeds $\simeq ~ c$ may have been formed.

Being the ultimate constituents of standard particles, superbradyons are expected to be also the constituents of the physical vacuum. They would play a fundamental dynamical and cosmological role, and could even lead to models of standard particle interactions without free Higgs-like bosons, somehow similar to the way the Cooper condensate generates the effective photon mass in superconductivity. 

Conventional particles can be the equivalent of phonons or solitons, as excitations of the superbradyonic ground state of matter (the vacuum) at the fundamental length scale. Their properties would then not be the ultimate fundamental properties of matter. Standard interactions may have been topologically generated in such a scenario. Also, our Universe may not be the only possible one. If different universes can also be formed, the properties of standard particles may be universe dependent. 

The aim of this note is to briefly discuss possible explanations of the electron and positron abundances reported by PAMELA \cite{PAMELA}, ATIC \cite{ATIC}, Fermi LAT \cite{FERMI}, HESS \cite{HESS} and PPB-BETS \cite{PPB-BETS}, in terms of superbradyonic physics. 

As present data do not clearly fit any specific dark matter model, it seems legitimate to explore unconventional scenarios related to all possible kinds of new physics. In the case of cosmic superbradyons, radiation effects at all energies seem {\it a priori} easier to observe than possible interactions with detectors. Furthermore, assuming that superbradyons are the exotic source of the observed high-energy electrons and positrons, this would not by itself imply that they are (most of) the gravitational dark matter. 

\section{Possible effects of cosmic superbradyons}

We assume the existence in our Universe of an absolute local rest frame (the vacuum rest frame, VRF), together with a fundamental length scale $a$ where new physics is expected to appear \cite{gonSL2}. A simple choice, but not compelling, would to identify $a$ with the Planck scale.

In our 1995 papers, the VRF was defined as the inertial frame where standard Lorentz symmetry for "ordinary" particles and the new space-time symmetry for the superluminal sector of matter were both exact symmetries up to very high-energy corrections due to the mixing between the two sectors. Actually, possible interactions between the superluminal particles and the physical vacuum of our Universe (similar, for instance, to refraction) should also be considered in such a description. In particular, superbradyons would interact with vacuum inhomogeneities.

The vacuum rest frame is expected to be close to that where the cosmic microwave background radiation appears to be isotropic. The addictiveness of energy and momentum for standard particles is assumed to be preserved in all inertial frames. Standard Lorentz transformations can then be used, leading to frame-dependent versions of the deformed relativistic kinematics and, possibly, to singularities in the UHE region \cite{gonSL2,Gonzalez-Mestres2004,gonSL3}.

In what follows, all calculations are performed in the VRF. 

It must be noticed that, although our LSV models \cite{gonSL1,gonSL2} have been presented with simple phenomenological formulae, their theoretical formulation is much more general. The approximations made are a consequence of the energy domain considered, where energies are small as compared to Planck energy or to any other fundamental energy scale. Linearly deformed relativistic kinematics (LDRK,\cite{gonSL2}) appears to be excluded by data, although hybrid models presenting a transition in the energy dependence of the deformation at very high energy can also be considered \cite{gonSL3}.

For simplicity, we also assume that, if superbradyons exist as free particles (cosmological remnants) or quasiparticles with our physical vacuum, their critical speed does not fundamentally differ from the value it may have in the absence of this vacuum or with another vacuum. Similarly, we consider here a situation with only one superluminal sector of matter and one superluminal critical speed, but our papers have also envisaged more general scenarios \cite{gonSL2}. The question of whether other universes can exist with different physical vacua is an important one, but the subject will not be addressed here. 

Energy and momentum conservation is assumed in any case to be valid at the energy and momentum scales considered here, as a simple consequence of space-time translation invariance. The situation can be more involved at a fundamental energy scale associated to $a$ or to the Planck length. 

Furthermore, energy and momentum conservation in all frames is a basic operating postulate of the patterns considered in \cite{gonSL1,gonSL2} where energy and momentum are taken to be additive for pairs of free particles, and standard Lorentz transformations lead to a frame-dependent kinematics and standard Lorentz symmetry is preserved in the VRF. Possible deviations from this hypothesis will be briefly discussed below.

\subsection{Superbradyons as sources of "ordinary" particles}

The superbradyons considered here as possible electron and positron sources are assumed to have comparative low kinetic energies. Superbradyons with kinetic energies closer to Planck scale will in principle be able to release more energetic cosmic rays (including the UHECR region \cite{gonSL1,gonSL2}). 

In the VRF of our Universe, assuming globally a superluminal Lorentz invariance from superbradyonic physics with $c_s~\gg ~c$ as the critical speed, the energy $E$ and momentum $p$ of a free superbradyon with critical speed in vacuum $c_s$ , inertial mass $m$ and speed $v$ would be : 
\begin{eqnarray}
E~=~c_s~(p^2~+~m^2 ~c_s^2)^{1/2} \\
p~=~m~v~(1 ~-~v^2~c_s^{-2})^{-1/2}
\end{eqnarray}

For $v~\ll ~c_s$ , these equations become :
\begin{eqnarray}
E~\simeq ~m~c_s^2~+~m~v^2/2 \\
p~\simeq ~m~v
\end{eqnarray}

The kinetic energy $E_{kin}$ being in this case :
\begin{equation}
E_{kin}~\simeq ~m~v^2/2
\end{equation}

In particular, $E_{kin}~\gg ~p~c$ for $v ~\gg ~c$ . Superbradyons with $v ~> ~c$ are kinematically allowed to decay by emitting standard particles, although the lifetimes for such processes can be very long because of the weak couplings expected. These decays may still exist in the present Universe and play a cosmological role.

For $v$ high enough, a superbradyon would be able to decay by emitting a set of conventional particles with total momentum $p_T~\ll ~E_T ~c^{-1}$ where $E_T$ is the total energy of the emitted set of particles. Such an event may fake the decay of a conventional heavy particle or contain pairs of heavy particles of all kinds.

The simplest configuration for the emitted particles would be a back-to-back particle - antiparticle pair, potentially able to fake possible conventional dark matter signatures. 

The same situation may arise if the superbradyon decays into one or several lighter superbradyons plus a set of conventional particles, or if our physical vacuum presents properties similar to those of low-gap superconductors. 

It therefore follows that data from PAMELA \cite{PAMELA}, ATIC \cite{ATIC}, Fermi LAT \cite{FERMI}, HESS \cite{HESS} and PPB-BETS \cite{PPB-BETS} can possibly be interpreted in terms of natural properties of a cosmological sea of superbradyons that are still decaying through the emission of "Cherenkov" radiation in vacuum or release conventional particles for some other reason. Whether or not data on electrons do exhibit a bump between 300 GeV and 600 GeV \cite{Malyshev} does not change this basic conclusion.

Then, the absence of hadronic signal as well as the observed spectra of electrons and positrons can possibly be explained by specific superbradyon energy spectra, interactions and decay modes in the present Universe. 

Superbradyons producing the observed electron and positron fluxes can have rest energies far above the TeV region, but this would not be necessarily the case. "Light" superbradyons with rest energies of the same order as those of the decay events or lower can also be considered. As an example, a superbradyon with $c_s ~\approx ~10^6~c$ and $m ~\approx $ 1 eV $c^{-2}$ would have a rest energy $ \approx $ 1 TeV. If its kinetic energy is of the same order, most of it can be spent in radiation in the form of "ordinary" particles. A $ \approx $ 1 TeV kinetic energy is to be compared to $m~c ~ \approx $ 1 eV $ c^{-1}$ corresponding to a (minimal) kinetic energy $\approx $ 0.5 eV.

Another possibility would be to consider annihilations of pairs of "light" superbradyons as the possible origin of the above astrophysical data.

If such "light" superbradyons can indeed exist, producing them in an accelerator experiment would normally be excluded because of the weak couplings expected. Even in the most favorable exceptional situations, identifying such (rare) accelerator events would be far from obvious. Possibilities would exist :

- If the produced superbradyons can exhibit detectable electromagnetic interactions.

- If some of them are produced in a very short-lived excited or unstable state favoring the emission of conventional particles through the "Cherenkov" effect in vacuum. Then, time of flight may allow to identify the event.

\subsection{Superbradyons and gravitation}

The energy of a superbradyon with inertial mass $m$ and $v~\simeq ~c$ would be dominated by the rest energy : $E~\simeq ~m~c_s^2~\gg ~ m~c^2$ . But we do not expect its effective gravitational mass with respect to standard gravitation to be that large, as direct coupling between superbradyons and standard matter is normally assumed to be very weak. Even so, superbradyons may significantly contribute to gravitational cosmological effects, including dark energy. They can also provide \cite{gonSL2} alternatives to the inflaton model, the transition from a purely superbradyonic universe to the present one at the Planck scale playing a fundamental role. 

The possibility of a superbradyon era in the evolution of our Universe at temperatures above $\approx ~10^{28}$ K or even lower, including the Grand Unification scale, has been considered in \cite{gonSL4}. As pointed out in this paper, this can be combined with an increasing role of deformed standard gravitation, as conventional matter is generated from the initial superbradyonic universe.  

An important issue is whether remnant superbradyons are concentrated close to standard astrophysical sources. It seems difficulty to answer this question. 

In principle, superbradyons are expected to be weakly coupled to gravitation, and this coupling can be as weak as $~\approx ~c_s^{-2}~c^2$ times that of standard gravitation. However, as the inertial mass of a superbradyon is $c_s^{-2}$ times its rest energy, this weak gravitational coupling can be enough to lead in practice to a strong effective reaction of superbradyonic matter to conventional gravitational fields. Then, superbradyons can strongly feel gravitation but generate very weak gravitational forces as compared to their total energy.

It is therefore not excluded that superbradyons be concentrated around the astrophysical objects usually considered in high-energy astronomy. 

\subsection{Other properties}

Superbradyon decays, annihilations and other interactions emitting "ordinary" particles would also increase the total effective gravitational mass in our Universe. Superbradyons may even be a source of more conventional dark matter particles, including those presently considered to explain data on electron and positron abundances. They would also manifest themselves through Lorentz symmetry violation (like QDRK) in the basic physics of standard particles \cite{gonSL2}.

Even if conventional particles are assumed to be made of superbradyons, a simple calculation shows that they are in any case kinematically prevented from decaying into superbradyons, except if they are very close to rest in the VRF. Otherwise, the conventional particle would fail to provide the energy required by the total momentum. As in all cases one has $E ~\gg ~p~ c_s $ for a superbradyon, the "ordinary" particle should necessarily fulfill a similar requirement.

For the same reason, in a superbradyonic era of our Universe, the production of standard matter would most likely have occured through "Cherenkov" decays and inelastic collisions rather than by direct formation of bound states.

\section{Superbradyons and quantum mechanics}

If matter is actually made of superluminal preons, the question of the validity of quantum mechanics for these ultimate constituents or below the fundamental length scale $a$ should also be addressed, even if it seems very difficult to propose real experimental tests on this issue. 

If the actual free superbradyons in our Universe are just quasiparticles generated from a new kind of (superbradyonic) condensed matter (the vacuum), the properties of such quasiparticles may be nontrivial. In any case, it is possible that the original fundamental superbradyons do not obey conventional quantum mechanics. Then, the usual quantum-mechanical laws would be a limit of the fundamental laws of standard matter for $p~a~\ll ~1$ (in Planck units), while new physics would govern the region $p~a~\gg ~1$ where superbradyons are the only "elementary" particles. 

Just to give an example, a simple hypothesis on the possible changes beyond the $a$ scale can be a transition from the standard Planck constant $h$ to a different value of the Planck constant, similar to the change from the conventional critical speed $c$ to the superluminal critical speed $c_s$. But more radical modifications can also be considered, for both space-time and quantum properties at distance scales smaller than $a$ . 

\subsection{Is standard quantum mechanics a composite phenomenon ?}

Can quantum mechanics be obtained from the properties of a new kind of composite medium, by combining a wave approach with topological constraints ? Are the standard gauge bosons and quantum numbers generated by topology ? Is quantization an exact property at all energies, or just a low-energy limit ?

To illustrate the idea very roughly, assume a real and positive function $f (\vec{x})$ defined on the standard three-dimensional space and such that the integral of $f^2$ on this space is quantified. This quantization can possibly be expressed as the limit of an energy cost for configurations.

Then, the value $n$ of such an integral could be related to the number of composite scalar particles. Time evolution may then lead a configuration fulfilling this condition (the value of the space integral) at a time $t ~=~0$ towards a set of $n$ spatially separate configurations with local $f^2$ integrals very close to 1 , equivalent to $n$ free particles. Otherwise, the $n$ particles would form a bound state. 

The possibility to use two complex coordinates as in \cite{gonSL5} instead of four real space-time variables, in order to directly account for half-integer spins, also deserves consideration.

Recent and current work on the grounds and interpretation of quantum mechanics \cite{QM} may provide useful guidelines and ideas to imagine how standard particle physics can be generated from superbradyon dynamics.

Thus, standard relativity and quantum mechanics would simultaneously cease to be fundamental laws of Nature below the critical length $a$. This may be normal, as both principles were derived from experiments on standard matter. 

Another important question deserving further study is whether the composite origin of quantum mechanics at the Planck scale or at some other fundamental scale can lead to a detectable energy-dependent violation of quantum mechanics and to observable effects at UHECR energies. 

Furthermore, if quantization can be actually the limit of the expression of an energy cost at the fundamental scale, possible tracks of non-optimal configurations should be investigated.

\subsection{Are energy and momentum conserved at energies close to Planck scale ?}

The validity of energy and momentum conservation above the fundamental energy and momentum scale is likely to depend on how well our description of matter can incorporate possible local vacuum inhomogeneities and similar phenomena. 

For the practical purposes considered in \cite{Gonzalez-Mestres2008,Gonzalez-Mestres2009,gonSL1}, it seems reasonable to require at least energy and momentum conservation with a precision better than $\approx ~10^{-4}$ eV at UHECR energies ($\approx ~10^{20}$ eV) in the models used, if standard calculations are to be preserved. This is just a minimal requirement, for physics involving hadrons and nuclei as UHECR. 

Patterns of energy of momentum violation should also be such that they prevent the observed UHECR from disappearing through unwanted spontaneous decays and similar processes.  

Then, for instance, if the violation of energy conservation grows like $E^3$, $E$ being the energy scale (similar to the energy dependence of the deformation term in QDRK), it will remain comparatively small (less than $\approx ~10^{20}$ eV) at the Planck scale ($E~ \approx ~10^{28}$ eV). But a ($\approx ~10^{-8}$ effect in the violation of energy conservation can be large enough to potentially produce important effects deserving further investigation. At the Grand Unification scale ($E~ \approx ~10^{24}$ eV), the violation of energy conservation would be less than $\approx ~100$ MeV. It would become less than $\approx ~10^{-28}$ eV at $E~ \approx ~1$ TeV, and less than $\approx ~10^{-37}$ eV at $E~ \approx ~1$ GeV. It seems to follow from these figures that an experimental effect due to energy nonconservation for standard particles can only be detected at very high energy. 

Questions such as vacuum structure, composition and time evolution or vacuum inner temperature, should also be dealt with beyond conventional schemes in such an approach. If energy and momentum are not exactly conserved, spontaneous particle emission and absorption by vacuum, and energy nonconservation inside vacuum, also deserve consideration. 

\section{Conclusion}

The hypothesis that the superluminal particles conjectured in our previous papers \cite{gonSL1,gonSL2} are at the origin of electron and positron abundances considered as possible experimental signatures of dark matter, seems worth exploring irrespectively of whether or not the sources of electrons and positrons are linked to a significant component of the gravitational dark matter. Several fundamental physics issues are involved in the discussion of the superbradyon hypothesis. 

Besides Lorentz symmetry violation and the actual properties of possible superluminal preons, the ideas developed here concern the origin and the interpretation \cite{QM,EPR} of quantum mechanics and a possible energy-dependent violation of standard quantum mechanics.  

It must be noticed, however, that the explanation of the above data in terms of dark matter does not seem to be the only possible one. More conventional astrophysical interpretations have been considered in \cite{FERMI,altern}. The question of the nature and role of dark matter has been evoked even in connection with spacecrafts and rockets \cite{space}, but associating electron and positron sources to gravitational dark matter is not a trivial step. 

A detailed discussion of the basic physical questions raised in this note will be presented in a forthcoming paper. 

\section{Post Scriptum to version v4}            

Describing superbradyons as particles in a similar sense to those of the Standard Model, except for the critical speed in vacuum, is obviously just a phenomenological approximation for practical purposes. A possible guess, that we follow in this paper, is that superbradyons exhibit a particle-like behavior if they can exist in our Universe as free objects or as vacuum excitations (quasiparticles...). This would not necessarily be the case for all superbradyons, and the superbradyonic objects exhibiting these properties would not necessarily be the "elementary" ones. 
 
Version v4 was written in August 2009. A few months later, WMAP has published its 7-year data and analyses \cite{WMAP,WMAPpapers}. These data have already raised several important questions.

\subsection{Possible implications of WMAP data}
\vspace{2mm}

The 7-year WMAP data have led in particular to the suggestion by Gurzadyan and Penrose \cite{Gurzadyan1} that concentric circles with anomalously low temperature variance can be identified in the cosmic microwave background (CMB) radiation sky, providing evidence for violent pre-Big-Bang activity. See also \cite{Moss,Wehus} and \cite{Gurzadyan2,Gurzadyan3} for recent analyses concerning these data. According to Gurzadyan and Penrose, such signals may be due to black-hole encounters in a previous aeon from a scenario based on conformal cyclic cosmology \cite{Penrose1,Penrose2} where standard inflation is also questioned. 

As stressed before and in our papers since 1995 \cite{gonSL2,gonSL4,Gonzalez-MestresInvUniv,Gonzalez-MestresCRIS2010}, superbradyons can provide an alternative to inflation and be the ultimate constituents of matter. The standard Big Bang can then be replaced by a superbradyon era, the superbradyons being cosmic superluminal preons. Then, the actual fundamental length where LSV and the superbradyonic world originate would in principle be smaller than the Planck length. In such a scenario, signals like those possibly found by Gurzadyan and Penrose can be naturally generated by this new noncyclic cosmology. Thus, simultaneously to UHECR experiments, measurements of the microwave background radiation may provide signatures of a primordial superbradyonic world. The situation is similar, more generally, for patterns of LSV generated by new physics beyond Planck scale, the superbradyon hypothesis being basically an ansatz to represent what this new physics can possibly be.  

LSV in a superbradyonic scenario or in any other noncyclic cosmology without the standard Big Bang or with a pre-Big-Bang era is not the only possible violation of a conventional fundamental principle that can potentially be detected by UHCER or CMB experiments. A long term study of UHECR and of cosmic rays at lower energies can potentially test \cite{gonSL4,Gonzalez-MestresCRIS2010,Gonzalez-MestresPamir}: quantum mechanics, the uncertainty principle, energy and momentum conservation, the number of effective space-time dimensions, the validity of hamiltonian and lagrangian formalisms, postulates of cosmology, vacuum dynamics and particle propagation, quark and gluon confinement, the elementariness of particles... Determining and exploiting the potentialities of CMB measurements in these domains will equally be a long-term task.

\subsection{Superbradyon production by very high energy cosmic rays}
\vspace{2mm}

The possibility that very high energy cosmic rays produce superbradyonic objects in the atmosphere or when interacting in space can also be considered, especially in LSV models with energy thresholds \cite{gonSL3,Gonzalez-MestresCRIS2010,Gonzalez-MestresPamir}. A new version of LSV has been recently proposed by Anchordoqui et al. \cite{Landsberg} where the number of effective space dimensions decreases with the energy scale through scale thresholds. According to the authors, this new LSV pattern may be a candidate to explain the jet alignment possibly observed by Pamir and other experiments above $\sim ~ 10^{16}$ eV \cite{mountain,stratosph}. However, as shown in \cite{Gonzalez-MestresPamir}, missing transverse energy in cosmic-ray interaction jets above some energy scale ($\sim ~ 10^{16}$ eV having in mind Pamir data) can be due to the production of superbradyonic objects (waves, particles...) involving a small portion of the total energy, captured from the target energy during the interaction, and a negligible fraction of the incoming momentum. Such an energy capture (by vacuum ?) would naturally lead to elongated jets, and subsequent polarization effects inside vacuum and secondaries may contribute to planar jet alignment \cite{Gonzalez-MestresCRIS2010}. In such a scenario, $v_s ~ \sim ~ c$ and $c_s ~ \sim ~10^6$ $c$ yield $m_s~\sim ~ 10^{-3}$ eV $c^{-2}$, $p_s ~\sim ~ 10^{-3}$ eV $c^{-1}$ and kinetic energy $\sim ~ 10^{-3}$ eV where $v_s$, $m_s$ and $p_s$ are the speed, mass and momentum of the superbradyonic object \cite{Gonzalez-MestresPamir}. The choice $c_s ~ \sim ~10^6$ $c$ is just an ansatz similar to the ratio between the speed of light and the speed of sound.

UHECR can also present similar phenomena. Although the captured energy would in all cases be limited by the target energy, if the vacuum captures nearly all this energy, a fall of the effective cross-sections of the observable interactions of the cosmic rays with the atmosphere can become apparent. Such an effect may even fake the GZK cutoff in some cases. At a scale above the highest observed cosmic-ray energies, a fall of the cross-sections of cosmic rays with the atmosphere had already been predicted in QDRK \cite{gon97d}. 

After the version v4 of this paper was completed and our CRIS 2008 talk was published \cite{Gonzalez-Mestres2009}, the HiRes final results have been presented \cite{HiRes-final}, as well as new data and analyses from the Pierre Auger Collaboration \cite{AUGER0910,AugerData4}. Reference \cite{Gonzalez-MestresCRIS2010} updates our analysis of the present situation concerning data from UHECR experiments, the tests of special relativity in the GZK energy region and global prospects for UHECR tests of fundamental principles. If the positron flux observed by PAMELA and other experiments \cite{PAMELA} were due to "Cherenkov" emission in vacuum by superbradyons with kinetic energy $~\sim ~1$ TeV, $v_s$ slightly above $c$ and $c_s ~ \sim ~ 10^6$ $c$, the superbradyon rest energy would be $m_s ~c_s^2 ~\sim ~10^{24}$ eV. Some spectacular decays could then perhaps be detected by UHECR experiments.

\subsection{Quantum field theory, cosmological constant, Higgs...}
\vspace{2mm}

Is there really a cosmological constant problem in nowadays particle physics and cosmology?

The possibility to circumvent the problems related to ultra-violet divergences in quantum field theory and to the value of the cosmological constant was already foreseen for LSV with QDRK \cite{gonSL2,gonSL3} considering the role of the deformation, the possible underlying dynamics and the new approach to vacuum structure. Indeed, these difficulties seem to be mainly related to vices of standard formulations originally based on low-energy physics and perturbative strategies. The recent suggestion by Anchordoqui et al., incorporating LSV and a VRF associated to an energy dependence of the effective dimensionality of space, is also claimed to potentially cure ultraviolet field-theoretical divergences \cite{Landsberg} and solve the cosmological constant problem \cite{GarciaA}. Similarly, the superbradyon scenario is expected to naturally solve these problems by introducing a new fundamental structure of matter and a new space-time geometry at very small distances. In any case, a new approach to quantum field theory seems necessary at the present stage for very high energy interactions and vacuum structure. As in previous periods in the development of particle physics, condensed matter physics can be a useful guide to elaborate new concepts adapted to current open questions \cite{gonSL1,gonSL2,Gonzalez-MestresCRIS2010}. 

In a composite approach where the standard particles would be built from the internal dynamics of a superbradyonic medium \cite{gonSL1}, the standard gauge bosons do not need to originate from local symmetries in the conventional sense. Contrary to former constituent models of standard particles where preons where considered as mere building blocks \cite{preons}, superbradyons can form a real new phase of matter where conventional particles are collective excitations of a large medium. Then, at the origin, taking for simplicity a lattice picture, the conventional gauge interactions can be associated to nearest-neighbour couplings (local potentials) describing the interaction between different local "vibration" modes of the physical vacuum. The vector bosons carrying conventional gauge-field forces would then be generated if these nearest-neighbour couplings turn out to depend on position, time and direction due to the existence of propagating vacuum excitations (the standard particles) involving the same family of "vibration" modes. In such a picture, conventional symmetries and charges would not necessarily be fundamental properties: they appear basically as a low-energy view of conventional matter, possibly ignoring important dynamical differences between the original "vibration" modes of vacuum at the superbradyonic scale. 

In this kind scenario, gravitation would be a composite phenomenon and the coupling of the vacuum to the gravitational field would be radically different from standard cosmology. Therefore, the gravitational field  equations can undergo essential changes and naturally incorporate an effective value of the cosmological constant compatible with phenomenology. As conventional particles would be made of "vibrations" of the superbradyonic vacuum, the Higgs bosons do not need to be statically materialized as a condensate in this vacuum. The Higgs mechanism can be a reaction of vacuum to the presence of gauge bosons and of other standard particles, similar to the energy capture envisioned in \cite{Gonzalez-MestresPamir} or to the above described generation of actual gauge bosons. Only the output of the real-time process in the presence of conventional particles would be coupled to standard gravitation in the usual way. Similarly, a dynamical view of vacuum as a medium of basic ultimate constituents, replacing the standard static one in terms of condensates of conventional particles, should allow to naturally avoid all ultraviolet divergences usually encountered in perturbative quantum field theory and to automatically preserve unitarity.  

A more detailed discussion of these basic questions will be presented elsewere. 

\subsection{Superbradyons and LHC}
\vspace{2mm}

Although one should not in principle expect to find at LHC the superbradyons considered here, other superbradyons may perhaps be produced at an observable rate and with detectable signals \cite{gonSL2}. This could be the case, for instance, for superbradyons similar to those emitted through the energy capture by vacuum considered in \cite{Gonzalez-MestresPamir}, even if for this scenario an energy threshold around $\sim ~ 10^{16}$ eV was postulated. 

The possible mixing between superbradyonic objects and standard particles at LHC energies deserves further study \cite{gonSL6}, having in mind constraints from : i) present experimental low-energy bounds on LSV \cite{Lamoreaux} ; ii) very high-energy cosmic-ray data where the interaction properties of standard particles are observed. 

\subsection{Further considerations}
\vspace{2mm}

It must also be noticed that, like photons in condensed matter, free superbradyons may undergo refraction in the physical vacuum of our Universe. Free superbradyonic matter may also exist in the present Universe only as quasiparticles and other forms of vacuum excitations, or be confined inside vacuum and standard particles, or be able to quit and enter this Universe... \cite{Gonzalez-MestresCRIS2010}. Then, the kinematics and critical speed of superbradyons in our vacuum would not be the same as in an "absolute" vacuum, assuming the latter can exist. These scenarios also deserve further study, even if we have basically ignored them aiming to present a simple picture. As previously stressed, it is even not clear whether superbradyons can indeed be described as particles in the usual sense. Similarly, superbradyonic constituents of matter could also lead to families of composite objects different from the standard particles and with a different critical speed in vacuum. Cosmic-ray experiments can in principle search for such new particles. Patterns where $c_s$ is infinite or extremely large, or strongly dependent on the age of our Universe, should also be further investigated for cosmological applications.  

Superbradyons can also provide alternatives to eternal inflation patterns \cite{Aguirre,Hartle} where our observable Universe is assumed to be inside a bubble nucleated in a de Sitter background. In particular, a superbradyonic "pre-Universe" would be able to explain possible WMAP signatures \cite{Feeney} that could be attributed to eternal inflation and to collisions between bubbles. Our present vacuum would be a piece of superbradyonic condensed matter (or some other phase) inside a much larger superbradyonic world, and our standard particles, excitations of this possibly condensed matter. The apparent expansion of our Universe would then correspond to the nucleation of a specific phase of superbradyonic matter where our standard particles can exist.

\end{document}